\documentclass[11pt,a4paper]{article}
\usepackage[utf8]{inputenc}
\usepackage{braket}
\usepackage{amsmath}
\usepackage[english,russian]{babel}
\usepackage{graphicx}
\graphicspath{ {imageseps/} }
\usepackage{hyperref}
\usepackage{physics}
\hypersetup{
    colorlinks=true,
    linkcolor=blue,
    filecolor=magenta,
    urlcolor=cyan,
}
\usepackage{tikz-feynman}

\begin{document}

\selectlanguage{english}

\hfill ITEP-TH-17/17

\vspace{5mm}

\centerline{\Large \bf Dynamical Casimir effect and surprises with loop corrections to it}

\vspace{5mm}

\centerline{E. T. Akhmedov and S. O. Alexeev}

\begin{center}
{\it Institutskii per, 9, Moscow Institute of Physics and Technology, 141700, Dolgoprudny, Russia}
\end{center}

\begin{center}
{\it B. Cheremushkinskaya, 25, Institute for Theoretical and Experimental Physics, 117218, Moscow, Russia}
\end{center}

\vspace{3mm}



\centerline{\bf Abstract}

We calculate quantum loop corrections to the stress--energy flux caused by moving mirrors. We consider massless, self--interacting, $\phi^4$, real scalar theory. In these calculations we encounter a new and quite unexpected subtleties due to the absence of global hyperbolicity in the presence of mirrors. We attempt to clearly phrase as many as possible hidden assumptions and complications that appear along the course of the solution of the problem in question. On top of that we find that quantum loop corrections to the stress--energy flux are growing with time and are not suppressed in comparison with the semi--classical contributions. Thus, we observe the break down of the perturbation theory, discuss its physical origin and ways to deal with such a situation. As a byproduct we observe a similarity of the problem in question with that for the minimally coupled, massless scalar field in de Sitter space.

\vspace{10mm}

\section{Introduction}

It is a commonly accepted opinion that the flux which is generated by a moving mirror is saturated by its semi--classical value \cite{BirDav}, \cite{FullDav}. The goal of the present paper is to show that this is not the case in a self--interacting theory. In fact, it is known in condensed matter theory that in non--stationary situations in non--Gaussian theories semi--classical approximation breaks down (see e.g. \cite{LL} and \cite{Kamenev}). The same situation is observed in de Sitter space quantum field theory \cite{AkhmedovKEQ}--\cite{Akhmedov:2013vka}), in the scalar QED on the strong electric field backgrounds \cite{Akhmedov:2014hfa}, \cite{Akhmedov:2014doa} and in the quantum corrections to the Hawking radiation \cite{AkhGodPop}.

We consider the massless, two--dimensional, real scalar field theory,

\begin{equation}\label{eq1}
S = \int_M d^2x \, \left[\partial_\mu \phi \, \partial^\mu \phi - \frac{\lambda}{4} \, \phi^4\right],
\end{equation}
in the presence of boundaries (mirrors) performing various types of motion. Namely, we assume that the space--time manifold $M$ has a time--like boundary and the boundary terms in the action are trivial.

To quantize the theory one represents the field operator $\phi(t,x)$ in the following form:

\begin{equation}\label{eq2}
\phi\left(\underline{x}\right)=\int_{0}^{\infty} \frac{dk}{2\pi} \big[a_k h_k(u,v) + a_k^+ \, \overline{h}_k(u,v)\big].
\end{equation}
The modes $h_k(u,v)$, where $\underline{x} = (t,x)$, $v = t+x$ and $u = t - x$, solve the Klein-Gordon (KG) equation, $\partial_\mu \partial^\mu \phi = 0$, and satisfy the following boundary conditions: $h_k\big[t \pm z(t)\big]=0,$ where $\left[t,z(t)\right]$ is a time-like curve (world-line of the mirror). The field operator $\phi(t,x)$ also obeys the same boundary condition $\phi[t, \, z(t)] = 0$.

In the rest of the Introduction section we describe the setup for the problem and unexpected subtleties that one encounter in the quantization of fields in the presence of perfect mirrors --- such mirrors which reflect modes with all possible momenta $k$ equally well. In the subsection 1.3 we show the mirror world--lines and mode functions. In the subsections 1.2 we continue with the calculation of the commutation relations in the presence of mirrors and describe the corresponding subtleties. The subsection 1.3 contains the discussion of the free Hamiltonian in the presence of mirrors and also related subtleties. In the subsection 1.4 we repeat the standard calculation of the semi--classical contribution to the stress--energy flux and concisely describe the reason for the relevance of the quantum loop corrections to it. Subsection 1.5 contains the calculation of the tree--level behavior of the Wightman two--point function at coincident points in the presence of mirrors. Here we point out the analogy of the present problem with the one for the massless minimally coupled scalar field in de Sitter space.

In the sections 2, 3 and 4 we calculate the two--loop sunset diagram corrections to the Keldysh propagator. In the section 5 we calculate the one loop bubble diagram corrections to the same propagator. We show that they grow with time, which signals the break down of the semi--classical approximation and of the perturbation theory. We discuss the consequences of such a secular growth to the stress--energy flux. Here again we observe an analogy between the behavior of massless two--dimensional scalars in the presence of mirrors and four--dimensional scalars in de Sitter space. Section 6 contains conclusions and discussion.

\subsection{Mirror world--lines and mode functions}

\begin{figure}[h]
\centering
\includegraphics[scale=0.25]{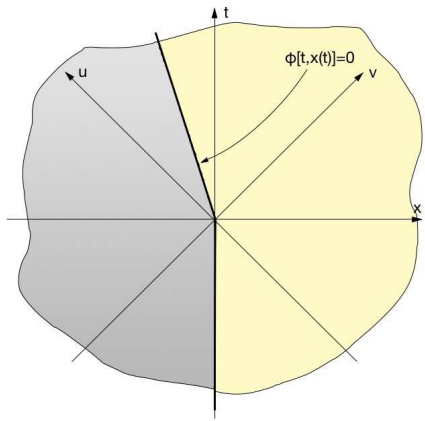}
\caption{The broken world-line. Falling towards the mirror wave is $e^{- i k v}$. To construct the mode functions one has to add to the falling wave different reflected ones in the regions $u>0$ and $u<0$.}
\label{fig:3}
\end{figure}

Most of the equations in this paper are applicable to generic mirror world--lines, but below we will be mainly interested in the following concrete situations. The first one is when the mirror stays at rest until $t=0$ and then instantaneously starts its motion with a constant speed, $\beta \leq 1$. The corresponding world--line is as follows (see fig. \ref{fig:3}):

\begin{equation}
\label{eq:6}
z(t)=\begin{cases}
\ \ 0, & t<0\\

  -\beta t, & t\ge0.
\end{cases}
\end{equation}
This world--line is not quite physically meaningfull, but we use it for illustrative reasons.

To find the modes in such a situation in different parts of the space-time one has to add different reflected waves
to the one which is falling towards the mirror, $e^{- i \, k \, v}$:

\begin{equation}
\label{eq:7}
h_k(u,v) = \frac{i}{\sqrt{2k}}\big[e^{-ikv}-\theta(-u)e^{-ik u}-\theta(u)e^{-ik \alpha u}\big],
\end{equation}
where $\alpha = \frac{1 - \beta}{1 + \beta}$. The normalization factor in this equation will be explained in the next subsection.

\begin{figure}[h]
\centering
\includegraphics[scale=0.25]{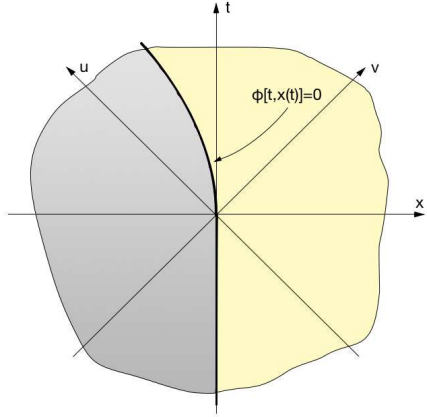}
\caption{The world-line of the mirror, approaching the speed $\beta \leq 1$.}
\label{fig:5}
\end{figure}

The second type of the mirror motion, that is interesting for us in the present paper, is described by the following world-line (see fig. \ref{fig:5}):

\begin{eqnarray}\label{worldline}
z(t)=\begin{cases}
\ \ 0, & t<0\\

  -\beta t+a(1-e^{-\beta t/a}), & t\ge0.
\end{cases}\end{eqnarray}
This world--line describes a mirror which homogeneously approaches, as $t \to + \infty$, a velocity which is less or equal to the speed of light, $\beta \leq 1$. If $\beta = 1$, then, as $t\to +\infty$, $1/a$ is proportional to the modulus of the four--acceleration and the world--line describes once started eternal accelerative motion.

The mode functions for an arbitrary mirror world-line $z(t)$ have the following form (see \cite{BirDav},\cite{FullDav}):

\begin{equation}
\label{eq:8}
h_k(u,v)=\frac{i}{\sqrt{2k}}\big[e^{-ikv}-e^{-ik(2t_u-u)}\big],
\end{equation}
where $t_u$ is a solution of the equation: $t_u-z(t_u)=u$. In fact, it is not hard to check that such a $h_k(u,v)$ solves the KG equation $\partial_u \partial_v \, h_k = 0$, because $t_u$ is a function of $u$ only, and obeys the corresponding boundary conditions.

\subsection{Commutation relations}

A quantum field theory on the Minkowskian half--plane may have problems due to the absence global hyperbolicity caused by the presence of a perfect mirror. Hence, we have to make two obvious checks: we have to find the form of the free Hamiltonian, which we do below, and to calculate the commutation relations between the field operator $\phi$ from eq.(\ref{eq2}) and its conjugate momentum $\pi=\partial_t \phi$ using the above defined modes (\ref{eq:8}). That is necessary because we have to check that $a_k$ and $a^+_k$ obey the appropriate commutation relations, i.e. they are actually the proper annihilation and creation operators.

Defining $\alpha(u) = \frac{1-\beta(t_u)}{1+\beta(t_u)}$, where $z'(t)=-\beta(t)$ and $1 \geq \beta(t) \geq 0$, we obtain:

\begin{eqnarray}\label{commut}
[\phi (t,x),\ \partial_t \phi(t,y)] = i\delta(x-y) - \nonumber \\
- \frac{i}{2} \bigg\{\alpha(u_y)\,\delta\big[v_x-2t(u_y)+u_y\big] + \delta\big[2t(u_x)-u_x-v_y\big]\bigg\}.
\end{eqnarray}
The last two terms on the r.h.s. do not vanish only when $x=y=z(t)$, i.e. when both points, $x$ and $y$, simultaneously lie on the boundary. So, in the region to the right of the boundary, we obtain the canonical commutation relations, $[\phi(t,x),\pi(t,y)]=i\delta(x-y)$, if  $a_k$ and $a^+_k$ obey the standard Heisenberg algebra.

For the broken world--line the calculation of the commutation relations is similar and straightforward. In that case we also encounter the boundary contributions to the commutation relations. Such contributions are present even in the case of standing mirror and in the case of the mirror moving with constant velocity, as can be also seen from (\ref{commut}).

Dropping off the boundary terms in eq. (\ref{commut}) is a subtle issue. Hence, it is worth discussing here their origin.
The non--canonical commutation relations appear due to the presence of the perfect mirror. In fact, that, vaguely speaking, leads to the reduction by half of the number of the degrees of freedom: so to say instead of $\cos(kx)$ and $\sin(kx)$ simultaneously we now have only $\sin(kx)$ and, furthermore, $k \in [0, +\infty)$ rather than $k\in (-\infty, +\infty)$. Rephrasing this, the presence of the boundary terms in the commutators is due to the fact that modes from eq. (\ref{eq:8}) are not related via a canonical transformation to those in empty space.

On top of that the modes under consideration have incorrect UV properties. That reveals itself e.g. via the presence of the second peculiarity in the commutator $\left[\phi(x), \, \phi(y)\right]$: besides the standard singularity appearing when $x$ is sitting on the light--cone emanating from $y$, there is a singularity in the commutator when $x$ is sitting on the light--cone emanating from $\bar{y}$ --- on the mirror image of $y$. This in particular means that $\left[\phi(x), \, \phi(y)\right]$ can be non--zero even if $x$ and $y$ are spatially separated. As a result of such a situation there are unusual UV divergences in the loops, which are not present in the standard (globally hyperbolic) situations\footnote{Note that in Euclidian signature $\bar{y}$ is sitting beyond the region where $x$ can take values. Hence, during the calculation of the loops in Euclidian signature one will encounter the standard UV divergences --- those which appear only when $x=y$. Thus, the results of the loop calculations in Euclidian signature do not coincide with those in Minkowskian one. That is true even in the stationary situations. It goes without saying that one should not expect any coincidence of the results of the loop calculations in Minkowskain and Euclidian signatures in non--stationary situations. Similar troubles one may encounter in the loops in the anti--de Sitter space vs. Lobachevsky constant negative curvature space of Euclidian signature \cite{Akhmedov:2012hk}.}.

One way to preserve the correct commutation relations and the proper UV behavior of the correlation functions is to add e.g. the following potential $\alpha \, \iint dt\, dx \, \delta\left[x-z(t)\right] \, \phi^2(t,x)$ to the action (\ref{eq1}). Then the equation for modes will be as follows:

$$
\left\{\partial^\mu \partial_\mu^{\phantom{\frac12}} + \,\, \alpha \, \delta\left[x - z(t)\right]\right\} \, \phi = 0
$$
and the $\delta$--functional potential would play the role of the reflecting barrier --- the mirror.

However, this is a subject of a separate study. In this paper we restrict our attention to the standard view of what is mirror, which is adopted e.g. in \cite{BirDav}. We try to clearly phrase all places where one can get into trouble during quantization in the presence of the standard mirror. In any case, the infrared secular effects that we consider below have nothing to do with the UV problems that we have described here.

\subsection{The free Hamiltonian}

In this subsection we find the free Hamiltonian for various mirror world-lines and show that even for non-stationary situations it is hermitian, but cannot be diagonalized once and forever. Nevertheless, some linear combination of the Hamiltonian $H$ and the momentum $P$, which is an evolution operator along the direction of the mirror world--line, can be diagonalized when the motion of the mirror becomes stationary. But in different regions of space--time such a diagonalization is performed by different types of modes, if the motion of the mirror is not stationary. That is the signal for the presence of a non--trivial stress--energy flux and for the secular growth of the loop corrections \cite{Akhmedov:2013vka}.

To illustrate the situation with the diagonalization of the free Hamiltonian we start with the consideration of the mirror moving with a constant velocity, $\beta \leq 1$. In this case the modes are as follows: $h_k(u,v)=\frac{i}{\sqrt{2k}}\big[e^{-ikv}-e^{-ik\alpha u}\big]$, where $\alpha = \frac{1-\beta}{1+\beta}$. Then it is straightforward to calculate that the free Hamiltonian depends on time:

\begin{eqnarray}\label{eq:hamiltonianconst}
H \equiv \frac{1}{2}\int_{0}^{\infty} dx \big[(\partial_t\phi)^2+(\partial_x\phi)^2\big] = \frac{1}{4}\int_{0}^{\infty}\frac{d\omega}{2\pi}(1+\alpha)\omega\bigg[a_\omega a_\omega^{\dag}+a_\omega^{\dag}a_\omega\bigg]+ \nonumber \\
+\frac{i}{2}(\alpha-1)\int_{0}^{\infty}\frac{d\omega}{2\pi}\int_{0}^{\infty}\frac{d\omega'}{2\pi}\frac{\sqrt{\omega\omega'}(\omega+\omega')}{(\omega+\omega')^2+0^2}\bigg[a_\omega a_{\omega'} e^{-it(1-\beta)(\omega+\omega')}-a_\omega^{\dag} a_{\omega'}^{\dag} e^{it(1-\beta)(\omega+\omega')}\bigg]+\nonumber \\
+\frac{i}{2}(\alpha-1)\int_{0}^{\infty}\frac{d\omega}{2\pi}\int_{0}^{\infty}\frac{d\omega'}{2\pi}\frac{\sqrt{\omega\omega'}(\omega-\omega')}{(\omega-\omega')^2+0^2}\bigg[ a_\omega a_{\omega'}^{\dag} e^{-it(1-\beta)(\omega-\omega')}-a_\omega^{\dag} a_{\omega'}e^{it(1-\beta)(\omega-\omega')}\bigg].
\end{eqnarray}
It means that it cannot be diagonalized once and forever, even if $\beta = const \leq 1$.
(At the same time, if $\beta = 1$, i.e. the mirror moves with the speed of light, the free Hamiltonian, $H$, is time independent and is diagonal.) This seems to be a very strange situation, because when the free Hamiltonian cannot be diagonalized once and forever that is a signal for the presence of a particle creation. However, the motion with a constant velocity is stationary and cannot lead to any flux. Thus, we seem to have here an apparent paradox.

But, at the same time, the momentum operator has a similar structure:

\begin{eqnarray}\label{eq:momentumconst}
P=\frac{1}{2}\int_{-\beta t}^{\infty} dx\ \big[\partial_t\phi\ \partial_x\phi+\partial_x\phi\ \partial_t\phi\big]= \frac{1}{4}\int_{0}^{\infty}\frac{d\omega}{2\pi}(1-\alpha)\omega\bigg[a_\omega a_\omega^{\dag}+a_\omega^{\dag}a_\omega\bigg]+ \nonumber \\
- \frac{i}{2}(\alpha+1) \int_{0}^{\infty}\frac{d\omega}{2\pi} \int_{0}^{\infty}\frac{d\omega'}{2\pi}\frac{\sqrt{\omega\omega'}(\omega+\omega')}{(\omega+\omega')^2+0^2} \, \bigg[a_\omega a_{\omega'} e^{-it(1-\beta)(\omega+\omega')}-a_\omega^{\dag} a_{\omega'}^{\dag} e^{it(1-\beta)(\omega+\omega')}\bigg] + \nonumber \\
-\frac{i}{2}(\alpha+1) \int_{0}^{\infty}\frac{d\omega}{2\pi} \int_{0}^{\infty}\frac{d\omega'}{2\pi}\frac{\sqrt{\omega\omega'}(\omega-\omega')}{(\omega-\omega')^2+0^2} \bigg[ a_\omega a_{\omega'}^{\dag} e^{-it(1-\beta)(\omega-\omega')}-a_\omega^{\dag} a_{\omega'}e^{it(1-\beta)(\omega-\omega')}\bigg].
\end{eqnarray}
As a result, the following linear combination of $H$ and $P$ has a very simple form and is diagonal:

\begin{equation}
\label{eq:lincomb}
H-\beta P=\frac{1}{2}\int_{0}^{\infty}\frac{d\omega}{2\pi}(1-\beta)\omega\bigg[a_\omega a_\omega^{\dag}+a_\omega^{\dag}a_\omega\bigg].\end{equation}
This result is not surprising, because the operator $H-\beta P$ defines translations along the world--line of the mirror.

For the case of the broken world--line, it is possible to represent the modes (\ref{eq:7}) in the following form:
$h_k(u,v)= \theta(u)h_k^\beta(u,v)+\theta(-u)h_k^0(u,v),$ where

\begin{eqnarray} \label{h0hbeta}
h_k^0(u,v)=\frac{i}{\sqrt{2k}}\big[e^{-ikv}-e^{-iku}\big],\quad {\rm and} \quad  h_k^\beta(u,v)=\frac{i}{\sqrt{2k}}\big[e^{-ikv}-e^{-ik\alpha u}\big].
\end{eqnarray}
Similarly one can represent the derivatives of the modes: $\partial_t h_\omega(u,v)=\theta(u)\partial_t h_\omega^{\beta}+\theta(-u)\partial_t h_\omega^{0},$ and $\partial_x h_\omega(u,v)=\theta(u)\partial_x h_\omega^{\beta}+\theta(-u)\partial_x h_\omega^{0}.$
Furthermore, the field operator $\phi$ and its derivatives can be rewritten in the same form. As a result, the linear combination $H - \beta(t) P$ is as follows:

\begin{equation} \label{eq:lincomb1}
\begin{split}
& H-\beta(t) P  =\theta(-t)H^0+\theta(t)\iint\frac{d\omega}{2\pi}\frac{d\omega'}{2\pi}\frac{\sqrt{\omega\omega'}}{2i}(\alpha-1)(1+\beta)\bigg[\pv \frac{1}{\omega+\omega'}\ a_\omega a_{\omega'}+ \\
& +\pv\frac{1}{-\omega-\omega'}\ a_\omega^{\dag} a_{\omega'}^{\dag}+\pv\frac{1}{\omega-\omega'}\ a_\omega a_{\omega'}^\dag+\pv\frac{1}{-\omega+\omega'}\ a_\omega^{\dag} a_{\omega'}\bigg]+\theta(t) \frac{1}{2}\int\frac{d\omega}{2\pi}\omega\big[a_\omega^{\dag}a_\omega+a_\omega a_\omega^{\dag}\big],
\end{split}
\end{equation}
where $H_0$ is the Hamiltonian operator for the mirror at rest and $\pv$ stands for the ``principal value''.

Unlike the case of the mirror moving with constant velocity, the linear combination $H - \beta(t) P$ cannot be diagonalized once and forever due to its explicit time dependence via $\theta(t)$. However before and after $t=0$, it is time independent and can be diagonalized by different basis of modes. Furthermore, from the above discussion we already know the behavior of the mode functions which do diagonalize the operator in question in the corresponding parts of space--time. They can be related to each other via a Lorentz boost.

For the case of generic motion, such as e.g. (\ref{worldline}), the Hamiltonian, $H$, and momentum, $P$, operators are obviously time dependent. However, the equations showing that, although being straightforward to obtain, are quite humongous, while the result is quite obvious, if one understands what is going on in the case of the broken world--line: When the motion becomes stationary $H$ and $P$ operators become time independent; If $0 < \beta < 1$, then as $t\to \pm \infty$ one can diagonalize only the linear combination $H - \beta \, P$; At the same time, if $\beta \to 0$ or $\beta \to 1$, as $t \to \pm \infty$, one can diagonalize $H$ and $P$ separately.

\subsection{Stress--energy flux}

The main goal of our paper is to calculate the quantum loop corrections to the stress--energy fluxes in the self--interacting field theory (\ref{eq1}) in the presence of mirrors performing various types of non--stationary motions. For the beginning, to set up the notations, we reproduce the well known formula for the semi--classical (tree--level) stress--energy flux due to a general mirror motion \cite{BirDav}. Then we show the results for the flux for some concrete world--lines that are described above. In the main body of the text we calculate loop corrections.

To calculate the expectation value of $T_{tx}$, we use the point--splitting regularization \cite{FullDav}:

\begin{equation}
\label{eq:ttxcommon}
\braket{T_{tx}}=\lim_{\varepsilon\to 0}\  \frac{1}{2}\left\langle \partial_t\phi(t,x)\partial_x\phi(t+i\varepsilon,x)^{\phantom{\frac12}} + \,\, \partial_x\phi(t,x)\partial_t\phi(t+i\varepsilon,x)\right\rangle,\end{equation}
where the average is taken with respect to the ground state $\ket{0}$, corresponding to the modes under consideration, $a_k |0\rangle = 0$, for all $k$. Substituting the expression of the field operator $\phi(t,x)$ through the modes $h_k(u,v)$ into the last expression, we obtain:

$$\braket{T_{tx}}=\lim_{\varepsilon\to 0}\  \frac{1}{4\pi}\int_{0}^{\infty} dk\ k\big[e^{-k\epsilon}-p'(u)p'(u+i\varepsilon)e^{ik[p(u+i\varepsilon)-p(u)]}\big],$$
where $p(u)=2 t_u-u$, $\displaystyle p'(u)=\frac{dp}{du}$ and $t_u$ is defined below the eq.(\ref{eq:8}).

Performing the integration and taking the limit $\epsilon \to 0$, we find that:

\begin{equation}\braket{T_{tx}} (u)=\frac{1}{24\pi}\bigg[\frac{p^{'''}}{p'}-\frac{3}{2}\bigg(\frac{p^{''}}{p^{'}}\bigg)^2\bigg]=-\frac{1}{12\pi}\  \frac{(1+v)^{1/2}}{(1-v)^{3/2}}\  \frac{d}{dt}\frac{\dot{v}}{(1-v^2)^{3/2}}\  \bigg|_{t=t_u},\end{equation}
where $v = \dot{z}$. As one can see, the result is expressed in terms of the mirror velocity and acceleration \cite{FullDav}. In fact, $\displaystyle \frac{\dot{v}}{(1-v^2)^{3/2}}$ is the mirror acceleration in its instantaneous rest frame\footnote{At the same time the result is expressed via the Schwarzian derivative of the Minkwoskian variant of the conformal transformation $u \to p(u)$, which is due to the conformal invariance of the theory under consideration. The theory (\ref{eq1}) is conformaly invariant only when $\lambda = 0$. In this paper we are interested in the case when $\lambda \neq 0$. Hence, we do not use the conformal invariance anywhere in our considerations below.}.

If the mirror moves with a constant velocity, then its proper acceleration is equal to zero, and the above formula gives $\left\langle T_{tx}\right\rangle =0$. In the case of the broken world-line the equation under consideration does not work, because the derivative of $p(u)$ is not defined. Nevertheless, we can easily obtain $\braket{T_{tx}}$ from the direct calculation:

\begin{eqnarray}\label{eq16}
\braket{T_{tx}} = \frac{1}{4\pi} \lim_{\varepsilon\to 0}\bigg\{\int_{0}^{\infty} k dk\ \bigg[  \theta(u)\theta(u+i\varepsilon)\big(e^{-k\varepsilon}-\alpha^2e^{-k\alpha\varepsilon}\big) + \theta(-u)\theta(-u-i\varepsilon)\big(e^{-k\varepsilon}-e^{-k\varepsilon}\big)\bigg] + \nonumber \\ + \int_{0}^{\infty} k dk\ \bigg[  \theta(u)\theta(-u-i\varepsilon)\big(e^{-k\varepsilon}-\alpha e^{-ik\alpha u + ik u - k\varepsilon}\big) + \nonumber \\+ \theta(-u)\theta(u+i\varepsilon)\big(e^{-k\varepsilon}-\alpha e^{-ik u + ik\alpha u -k\alpha\varepsilon}\big)\bigg]\bigg\} = \nonumber \\
= \frac{1}{8\pi}\delta'(u)\big[\theta(u) - \theta(-u)\big].
\end{eqnarray}
Despite the fact that this result is defined only as a generalized function, it meets the physical requirements, according to which the flux is not equal to $0$ only when $u=0$, i.e. when the motion of the mirror is not stationary.

For the world-line, which approaches the speed of light, as $t\to + \infty$ (see fig. \ref{fig:5} and eq. (\ref{worldline}) when $\beta = 1$), using the above defined general formula, we obtain:

$$
\braket{T_{tx}}=\frac{1}{12\pi a^2}\frac{2e^{-t_u/a}-1}{(e^{-t_u/a}-2)^4}.
$$
When $t$ goes to infinity, the energy flux becomes time independent and equal to $\displaystyle -\frac{1}{2^4 \cdot12\pi a^2}.$ That is because the motion of the mirror becomes stationary, as $t \to + \infty$. Similar situation appears in \cite{AkhGodPop} in the case of the thin shell collapse, which is a well established fact in the literature \cite{BirDav}, \cite{FullDav}.

To calculate the expectation value of the stress--energy flux we use the so called Keldysh propagator $G^K$:

\begin{eqnarray} \label{KeldyshSE}
\left\langle T_{tx}\right\rangle = \left.\frac12 \partial_{x_0} \partial_{y_1} G^K\left(\underline{x}, \, \underline{y}\right) \right|_{\underline{x} = \underline{y}}, \quad {\rm where} \quad G^K\left(\underline{x}, \, \underline{y}\right) \equiv \frac12 \, \left\langle \left\{\phi\left(\underline{x}\right)^{\phantom{\frac12}}, \, \, \phi\left(\underline{y}\right)\right\}\right\rangle
\end{eqnarray}
and $\{,\}$ is the anti--commutator.

Hence, the flux appearing from the tree--level propagator is just due to the amplification of the zero--point fluctuations: In other words, the flux is due to the fact that $h_k(u,v)$ is not just equal to a simple exponential function, as can be seen e.g. from eq.(\ref{eq16}).

However, that is not the whole story in the full self--interacting, e.g. $\lambda \, \phi^4$, theory. The point is that in the Gaussian approximation the creation and annihilation operators, $a_k$ and $a_k^+$, do not depend on time --- the whole time dependence of these operators is in the modes $h_k$. But if one turns on self--interactions the creation and annihilation operators start to depend on time. As a result, the quantum averages, $\left\langle a_k^+ \, a_{k'}\right\rangle$ and $\left\langle a_k a_{k'}\right\rangle$, also change in time, i.e. they can become non--zero even if they were vanishing at the past infinity. This dependence reveals itself in the behavior of the Keldysh propagator:

\begin{eqnarray}\label{eq14}
G^K\left(\underline{x},\underline{y}\right) = \iint \frac{dk}{2\pi}\frac{dk'}{2\pi} \left[\left(\frac{\delta\left(k-k'\right)}{2} + \left\langle a_k^+ \, a_{k'}\right\rangle\right) \, \bar{h}_k(\underline{x}) h_{k'}(\underline{y}) + \left\langle a_k a_{k'}\right\rangle \, h_k(\underline{x}) h_{k'}(\underline{y})+\operatorname{h.c.}\right].
\end{eqnarray}
In generic situations $\left\langle a_k^+ \, a_{k'}\right\rangle$ and $\left\langle a_k a_{k'}\right\rangle$ grow with time, if one calculates them in the approximation of the first few loops. That is the physical origin of the secular growth that we are studying in the present paper. Such a growth is practically inevitable at every order of the perturbation theory \cite{LL}, \cite{Kamenev}: In fact, there will {\it not} be any secular growth in the loops only if modes are plane waves, $\left\langle a_k a_{k'}\right\rangle = 0$ and $\left\langle a_k^+ \, a_{k'}\right\rangle = n_k \, \delta\left(k-k'\right)$, where $n_k$ is the Planckian distribution. None of the above conditions is true in the presence of the mirrors that perform an non--stationary motion.

The presence of the secular growth of the propagator drastically changes the flux that is calculated at the tree--level. In first place it leads to the violation of the semi--classical approximation. The goal of the present paper is to explicitly show this fact and to trace its physical consequences.

It is worth stressing here that the presence of $\left\langle a_k^+ \, a_{k'}\right\rangle$ means that in the self--interacting theory there is an excitation of higher levels (for the exact modes) on top of the amplification of the zero--point fluctuations, while the presence of the anomalous average $\left\langle a_k a_{k'}\right\rangle$ signals a change of the ground state of the theory in question. The fact that $\left\langle a_k^+ \, a_{k'}\right\rangle$ and
$\left\langle a_k a_{k'}\right\rangle$ are not necessarily proportional to $\delta\left(k+k'\right)$ and/or to $\delta\left(k-k'\right)$ in the situations that we consider in the present paper is due to the violation of the Lorentz invariance caused by the presence of mirrors.

\subsection{The tree--level Wightman propagators at coincident points}

In this subsection we consider the tree-level behaviour of the Wightman two--point function at the coincident points for different mirror world--lines. For an arbitrary motion the result is as follows:

\begin{equation*}
\begin{split}
& G^W(x,y)\big|_{x\to y} \equiv \braket{\phi(t,x)\phi(t+i\epsilon,x)}= \int_{0}^{\infty} \frac{dk}{2\pi}\frac{1}{2k}\left[e^{-ikv}-e^{-ikp_u}\right]\left[e^{ik(v+i\epsilon)}-e^{ikp(u+i\epsilon)}\right] \approx\\
& \approx \frac{1}{4\pi} \log\frac{(p_u-v)^2}{\epsilon^2p'(u)},\quad {\rm as} \quad \epsilon \to 0.
\end{split}\end{equation*}
First, consider the mirror world--line (\ref{worldline}) for $\beta < 1$.
In such a case $p(u)$ and $p'(u)$ for $u\to \infty$ are as follows:

\begin{eqnarray}
p(u)=t_u(1-\beta)+a\big(1-e^{-\beta t_u/a}\big)\approx t_u(1-\beta)\approx \alpha u, \nonumber \\
{\rm and} \quad p'(u)=\frac{1-\beta(1- e^{-\beta t_u/a})}{1+\beta(1- e^{-\beta t_u/a})}\approx \frac{1-\beta}{1+\beta}\equiv\alpha.
\end{eqnarray}
As a result, the Wightman propagator at the coincident points is as follows:

\begin{equation}
G^W(x,y)\big|_{x\to y}\approx \frac{1}{4\pi}\log \frac{\big(\alpha u-v\big)^2}{\alpha\epsilon^2},\quad {\rm when} \quad u\to \infty.
\end{equation}
Second, consider the mirror world--line (\ref{worldline}) for $\beta = 1$. In this case along the same lines we obtain that the Wightman propagator grows linearly in $u$ and logarithmically in $v$:

\begin{equation}
G^W(x,y)\big|_{x\to y}\approx \frac{1}{4\pi}\log \frac{2e^{u/2a}(a-v)^2}{\epsilon^2}= \frac{1}{4\pi}\bigg[\frac{u}{2a}+\log\frac{2(a-v)^2}{\epsilon^2} \bigg],\quad {\rm as} \quad u\to \infty.
\end{equation}
Such a behavior of the Wightman function is very similar to the one of the massless minimally coupled scalar field in the four--dimensional de Sitter space--time: See e.g. \cite{Starobinsky} and also \cite{Tsamis:2005hd},  \cite{Gautier:2015pca}--\cite{Serreau:2013psa} for the explanations and extensions. This is the first sign favouring the consideration of the massless two--dimensional scalar fields in the presence of mirrors as a model example for the massless minimally coupled scalar fields in four--dimensional de Sitter space. Below we will see that the loop contributions in the case under study have also very similar behaviour to those in the case of the massless scalars in de Sitter space.

The similarity between the four--dimensional massless minimally coupled scalar field theory in de Sitter space and the flat space two--dimensional scalar field theory has been noticed already long ago and in many places. Please note that the Wightman propagators in both theories have the same logarithmic dependence on the distance (see e.g. \cite{Tsamis:2005hd}). The novel feature that we observe here is that the similarity can go beyond the tree--level, if one considers the massless scalar in two--dimensions in the presence of mirrors.

\section{Two--loop corrections to the propagators}

As was explained above, to calculate the expectation value of the stress--energy flux we use the Keldysh propagator (\ref{KeldyshSE}).
The goal of the rest of the paper is to show that the value of this propagator in generic situations is not saturated by its tree--level contribution and the loop corrections to it can be of the same order. That may drastically change the found above values for the stress--energy flux, as was already mentioned.

In non--stationary situations any field is characterised by the three independent propagators (see e.g. \cite{LL} and \cite{Kamenev}):

\begin{equation}
\begin{split}
\label{eq:throughmodes}
& G^{--}_{xy}=\braket{T\phi(\underline{x})\phi(\underline{y})}=\theta(x^0-y^0)\int \frac{dk}{2\pi}\ h_k(\underline{x})\bar{h}_k(\underline{y})+\theta(y^0-x^0)\int \frac{dk}{2\pi}\ h_k(\underline{y})\bar{h}_k(\underline{x}); \\
& G^{++}_{xy}=\braket{\tilde{T}\phi(\underline{x})\phi(\underline{y})}=\theta(x^0-y^0)\int \frac{dk}{2\pi}\ h_k(\underline{y}) \bar{h}_k(\underline{x}) + \theta(y^0-x^0)\int \frac{dk}{2\pi}\ h_k(\underline{x})\bar{h}_k(\underline{y}); \\
& G^{+-}_{xy}=\braket{\phi(\underline{x})\phi(\underline{y})} = \int \frac{dk}{2\pi}\ h_k(\underline{x})\bar{h}_k(\underline{y}), \quad {\rm and} \quad G^{-+}_{xy} = \braket{\phi(\underline{y}) \phi(\underline{x})}=G^{+-}_{yx}.
\end{split}
\end{equation}
After the Keldysh rotation \cite{LL}, \cite{Kamenev}, every field is characterized by the Keldysh propagator:

\begin{equation}
\label{eq:keld1}
G^{K}\left(\underline{x}, \, \underline{y}\right) = \frac{1}{2}\big[G^{--}_{xy} + G^{++}_{xy}\big] \equiv \frac12 \, \left\langle \left\{\phi(\underline{x})^{\phantom{\frac12}},\,\, \phi(\underline{y})\right\}\right\rangle,\end{equation}
and by the retarded and advanced propagators, which are proportional to the commutator $\left[\phi(x), \, \phi(y)\right]$. While at the tree--level the commutator, being a $c$--number, is not sensitive to the state with respect to which the quantum averaging is performed, the Keldysh propagator traces the destiny of the state during the time evolution (see e.g. \cite{LL}, \cite{Kamenev}). The latter fact e.g. can be seen from eq. (\ref{eq14}).

In this paper we are interested in the behaviour of the loop corrections to these propagators as $(x_0 + y_0)/2 \to + \infty$, when $x_0 - y_0 = const$. We would like to know the destiny of the correlation functions of the theory in the future infinity.

Due to their causal structure the retarded and advanced propagators do not receive growing with time quantum loop corrections \cite{Kamenev}, \cite{Akhmedov:2013vka} in the first few loops, if we consider the above defined limit. However, they do receive growing corrections in the limit $\left|x_0 - y_0\right|\to \infty$ already in the first few loops. But this is not of interest for us in the present paper, because the secular growth of this sort does not affect the stress--energy flux. In fact, the latter growth usually leads to a mass renormalization or to an imaginary contribution to the self--energy, which is signaling the quasi--particle instability. At the same time, the stress--energy flux is affected only when there is a growth in the correlation functions at the coincident points (see e.g. \cite{Akhmedov:2017ooy} for the related discussion).

Now, the Keldysh propagator does receive secularly growing corrections in the limit of interest for us.
We start with the calculation of the two--loop correction to the Keldysh propagator, which follows from the sunset diagrams:
\begin{equation}
\label{eq:diag}
\Delta G^{K}_{xy}=\sum_{\sigma,\sigma_i=\pm} \begin{gathered}
\begin{tikzpicture}
\begin{feynman}
 \vertex[label=above: ${x, \sigma}$] (x);
    \vertex[label=above: ${z, \sigma_1 }\quad \ \ \ $] [right=2cm of x] (z);
    \vertex[label=above: $\quad \ \ \ \ {w, \sigma_2}$] [right=2.5cm of z] (w) ;
    \vertex[label=above: ${y, \sigma}$] [right=2cm of w] (y);
    \diagram* {
      (x) -- [plain] (z) -- [plain] (w) -- [plain] (y)
      (z) -- [plain, out=80, in=100] (w)
      (z) -- [plain, out=-80, in=-100] (w) };
\end{feynman}
 \end{tikzpicture}
 \end{gathered}
\end{equation}
This is the shorthand notation of the following sum:

\begin{equation*}
\begin{split}
& \begin{gathered}
\begin{tikzpicture}
\begin{feynman}
 \vertex[label=above:${x, +}\quad \ $] (x);
    \vertex[label=above: ${z, +}\quad \ $] [right=1.cm of x] (z);
    \vertex[label=above: $\quad \ \ {w, +}$] [right=1.5cm of z] (w) ;
    \vertex[label=above: $\quad \ \ {y, +}$] [right=1cm of w] (y);
    \diagram* {
      (x) -- [plain] (z) -- [plain] (w) -- [plain] (y)
      (z) -- [plain, out=80, in=100] (w)
      (z) -- [plain, out=-80, in=-100] (w) };
\end{feynman}
 \end{tikzpicture}
 \end{gathered} +\begin{gathered}
 \begin{tikzpicture}
\begin{feynman}
 \vertex[label=above:${x, +}\quad \ $] (x);
    \vertex[label=above: ${z, +}\quad \ $] [right=1.cm of x] (z);
    \vertex[label=above: $\quad \ \ {w, -}$] [right=1.5cm of z] (w) ;
    \vertex[label=above: $\quad \ \ {y, +}$] [right=1cm of w] (y);
    \diagram* {
      (x) -- [plain] (z) -- [plain] (w) -- [plain] (y)
      (z) -- [plain, out=80, in=100] (w)
      (z) -- [plain, out=-80, in=-100] (w) };
\end{feynman}
 \end{tikzpicture} \end{gathered} + \begin{gathered}
\begin{tikzpicture}
\begin{feynman}
 \vertex[label=above:${x, +}\quad \ $] (x);
    \vertex[label=above: ${z, -}\quad \ $] [right=1.cm of x] (z);
    \vertex[label=above: $\quad \ \ {w, +}$] [right=1.5cm of z] (w) ;
    \vertex[label=above: $\quad \ \ {y, +}$] [right=1cm of w] (y);
    \diagram* {
      (x) -- [plain] (z) -- [plain] (w) -- [plain] (y)
      (z) -- [plain, out=80, in=100] (w)
      (z) -- [plain, out=-80, in=-100] (w) };
\end{feynman}
 \end{tikzpicture}
 \end{gathered} + \\
 + & \begin{gathered}
\begin{tikzpicture}
\begin{feynman}
 \vertex[label=above:${x, +}\quad \ $] (x);
    \vertex[label=above: ${z, -}\quad \ $] [right=1.cm of x] (z);
    \vertex[label=above: $\quad \ \ {w, -}$] [right=1.5cm of z] (w) ;
    \vertex[label=above: $\quad \ \ {y, +}$] [right=1cm of w] (y);
    \diagram* {
      (x) -- [plain] (z) -- [plain] (w) -- [plain] (y)
      (z) -- [plain, out=80, in=100] (w)
      (z) -- [plain, out=-80, in=-100] (w) };
\end{feynman}
 \end{tikzpicture}
 \end{gathered} +\begin{gathered}
 \begin{tikzpicture}
\begin{feynman}
 \vertex[label=above:${x, -}\quad \ $] (x);
    \vertex[label=above: ${z, -}\quad \ $] [right=1.cm of x] (z);
    \vertex[label=above: $\quad \ \ {w, -}$] [right=1.5cm of z] (w) ;
    \vertex[label=above: $\quad \ \ {y, -}$] [right=1cm of w] (y);
    \diagram* {
      (x) -- [plain] (z) -- [plain] (w) -- [plain] (y)
      (z) -- [plain, out=80, in=100] (w)
      (z) -- [plain, out=-80, in=-100] (w) };
\end{feynman}
 \end{tikzpicture} \end{gathered} + \begin{gathered}
\begin{tikzpicture}
\begin{feynman}
 \vertex[label=above:${x, -}\quad \ $] (x);
    \vertex[label=above: ${z, -}\quad \ $] [right=1.cm of x] (z);
    \vertex[label=above: $\quad \ \ {w, +}$] [right=1.5cm of z] (w) ;
    \vertex[label=above: $\quad \ \ {y, -}$] [right=1cm of w] (y);
    \diagram* {
      (x) -- [plain] (z) -- [plain] (w) -- [plain] (y)
      (z) -- [plain, out=80, in=100] (w)
      (z) -- [plain, out=-80, in=-100] (w) };
\end{feynman}
 \end{tikzpicture}
 \end{gathered} + \\
 + & \begin{gathered}
\begin{tikzpicture}
\begin{feynman}
 \vertex[label=above:${x, -}\quad \ $] (x);
    \vertex[label=above: ${z, +}\quad \ $] [right=1.cm of x] (z);
    \vertex[label=above: $\quad \ \ {w, -}$] [right=1.5cm of z] (w) ;
    \vertex[label=above: $\quad \ \ {y, -}$] [right=1cm of w] (y);
    \diagram* {
      (x) -- [plain] (z) -- [plain] (w) -- [plain] (y)
      (z) -- [plain, out=80, in=100] (w)
      (z) -- [plain, out=-80, in=-100] (w) };
\end{feynman}
 \end{tikzpicture}
 \end{gathered} +\begin{gathered}
 \begin{tikzpicture}
\begin{feynman}
 \vertex[label=above:${x, -}\quad \ $] (x);
    \vertex[label=above: ${z, +}\quad \ $] [right=1.cm of x] (z);
    \vertex[label=above: $\quad \ \ {w, +}$] [right=1.5cm of z] (w) ;
    \vertex[label=above: $\quad \ \ {y, -}$] [right=1cm of w] (y);
    \diagram* {
      (x) -- [plain] (z) -- [plain] (w) -- [plain] (y)
      (z) -- [plain, out=80, in=100] (w)
      (z) -- [plain, out=-80, in=-100] (w) };
\end{feynman}
 \end{tikzpicture} \end{gathered}
\end{split}
\end{equation*}
We will use such notations below.

The reason why we start our considerations with such two--loop contributions rather than with the one--loop bubble corrections is because even if the bubble diagrams do grow with time their contribution just leads to the (time--dependent) mass renormalization and does not affect the stress--energy flux. We discuss one--loop diagrams in the last two sections.

In the limit $(x^0+y^0)/2\to \infty$, when $|x^0-y^0|=\operatorname{const}$, the leading contributions
to the Keldysh propagator {\it at any loop order} can be represented as follows:

\begin{equation}\label{GK}
\Delta G^{K}\left(\underline{x}, \, \underline{y}\right) =\iint \frac{dk}{2\pi}\frac{dk'}{2\pi} \big[n_{kk'} \bar{h}_k(\underline{x}) h_{k'}(\underline{y})+\kappa_{kk'} h_k(\underline{x}) h_{k'}(\underline{y}) + \operatorname{h.c.}\big].
\end{equation}
A simple way to see the origin of this equation is to notice eq. (\ref{eq14}) and to observe that $n_{kk'} = \left\langle 0 \left| T_C\left[S \, a^+_k \, a_{k'}\right]\right| 0 \right\rangle$ and $\kappa_{kk'} = \left\langle 0 \left| T_C\left[S \, a^+_k \, a_{k'}\right]\right| 0 \right\rangle$, where $T_C$ is the ordering along the Keldysh time contour \cite{Kamenev}, $S$ is the evolution operator in which we adiabatically turn on self--interactions after $t_0$; and $a^+_k$, $a_k$ are placed at $x_0$, while $a^+_{k'}$ and $a_{k'}$ --- at $y_0$.

Thus, in general $n$ and $\kappa$ depend on both points $x_0$ and $y_0$. But, if we put $x^0\approx y^0\approx T\to \infty$ and keep track of only the leading contributions, the correction from the sunset diagram can be expressed as follows:

\begin{equation}
\label{eq:nkk}
n_{kk'} (T) \approx 2\lambda^2\int d^2z\  d^2w\ \theta(T-w^0)\theta(T-z^0)\  {h}_k(z) \bar{h}_{k'}(w) \bigg[\int \frac{dp}{2\pi} h_p(z) \bar{h}_p(w)\bigg]^3,\end{equation}
and

\begin{equation}\label{eq:kappakk}
\kappa_{kk'}(T) \approx -2\lambda^2\int d^2z\  d^2w\ \theta(T-z^0)\theta(z^0-w^0)\   \bigg[\bar{h}_k(z) \bar{h}_{k'}(w)+\bar{h}_k(w) \bar{h}_{k'}(z)\bigg] \bigg[\int \frac{dp}{2\pi} h_p(z) \bar{h}_p(w)\bigg]^3.\end{equation}
Below we explicitly calculate the leading behaviour of $n_{kk'}(T)$ and $\kappa_{kk'}(T)$, as $T\to \infty$, via the substitution into the above expressions different concrete forms of the mode functions $h_k$ for various types of mirror motions. The physical meaning of $n$ and $\kappa$ is discussed in the subsection 1.4 above and in the concluding section.

\section{The two-loop corrections to $n_{kk'}$} \label{twoloopnkk} \

In this section we calculate $n_{kk'}$ from eq. (\ref{eq:nkk}) for various types of mirror world-lines.

\subsection{The situation without a mirror}

To clarify the physical meaning of the contributions that we consider here let us first consider the simplest case, when there is no mirror and we deal with the standard scalar field theory in the entire $1+1$ Minkowski space-time. For this calculation we keep the mass $m$ of the field arbitrary\footnote{Note that in the calculation of the loop contribution above we did not use anywhere the presence of the mirror and the fact that the field is massless. The equations (\ref{GK}), (\ref{eq:nkk}) and (\ref{eq:kappakk}) are valid for any 2D situation with the real scalar field.} for the reasons that become clear in the next subsection. Modes in this case are plane waves of the form:

$$h_k(t,x)=\frac{1}{\sqrt{2\varepsilon_k}}e^{-i(\varepsilon_k t-kx)},\  \operatorname{where}\ \varepsilon_k=\sqrt{k^2+m^2}\ \operatorname{and}\ -\infty<k<\infty.$$
Then the integral over the coordinates $z=(z^0, z^1)$ and $w=(w^0, w^1)$ in eq. (\ref{eq:nkk}) can be rewritten as follows:

\begin{equation}\int_{t_0}^T dz^0\ e^{-i(\varepsilon_k+\sum_i \varepsilon_{p_i})z^0} \int_{t_0}^T dw^0\   e^{i(\varepsilon_{k'}+\sum_i \varepsilon_{p_i})w^0}\int_{-\infty}^{\infty} dz^1\ e^{i(k+\sum_i p_i)z^1} \int_{-\infty}^{\infty} dw^1\ e^{-i(k'+\sum_i p_i)w^1}.\end{equation}
Here $t_0$ is  the moment of time after which we start to adiabatically turn on the self--interactions.

In the limit $T - t_0 \to +\infty$ this expression is proportional to

\begin{equation}\begin{split} & \delta\bigg(\varepsilon_k+\sum_i \varepsilon_{p_i}\bigg)\delta\bigg(\varepsilon_{k'}+\sum_i \varepsilon_{p_i}\bigg)\delta\bigg(k+\sum_i p_i\bigg)\delta\bigg(k'+\sum_i p_i\bigg) \propto \\
& \propto(T-t_0)\delta(k-k')\delta\bigg(\varepsilon_k+\sum_i \varepsilon_{p_i}\bigg)\delta\bigg(k+\sum_i p_i\bigg),\end{split}\end{equation}
where $\delta(0)\propto T-t_0$.

Thus, the r.h.s. of (\ref{eq:nkk}) is proportional to $T-t_0$, i.e. seemingly grows with time, but $\left(\varepsilon_k+\sum_i \varepsilon_{p_i}\right)$ is never equal to zero. Hence, $\delta\left(\varepsilon_k+\sum_i \varepsilon_{p_i}\right)$ and, correspondingly, the correction to $n_{kk'}$ are equal to zero.


The calculation above illustrates the fact that the secular growth in a stationary situation is generally forbidden by the energy conservation. However, if there is a violation of the energy conservation, such as the one present in the background field (if one does not include this field to close the whole system) or in the background of a non--stationary moving mirror, then $n_{kk'}$ would generically grow with time, even if it was zero at the beginning. We will see that below.

\subsection{The mirror at rest}

Now let us consider the case, when there is a mirror at rest. To calculate $n_{kk'}$ from eq.(\ref{eq:nkk}), let us first take the integral over the variable $z$:

\begin{equation}\begin{split} & \int d^2 z\  \theta(T-z^0)\  h_k(z) h_{p_1} (z) h_{p_2} (z) h_{p_3} (z) = \\
& =\frac{1}{\sqrt{2^4\cdot kp_1p_2p_3}} \int_{t_0}^{T} dz^0\  e^{-i(k+p_1+p_2+p_3)z^0} \int_{0}^{\infty} dz^1 \sum_{\sigma,\sigma_i=\pm 1} \sigma\sigma_1\sigma_2\sigma_3\  e^{-i(\sigma k +\sigma_1 p_1+\sigma_2 p_2 +\sigma_3 p_3)z^1}.
\end{split}
\end{equation}
This expression is proportional to:

\begin{equation*}
\frac{1}{\sqrt{k p_1 p_2 p_3}}\bigg(\sum_{\sigma,\sigma_i=\pm 1} \sigma\sigma_1\sigma_2\sigma_3\ \frac{1}{i(\sigma k+\sum_i \sigma_i p_i-i\epsilon)}\bigg) \int_{t_0}^{T} dz^0\  e^{-i(k+p_1+p_2+p_3)z^0}.
\end{equation*}
When $T - t_0 \to +\infty$, the last integral is approximately equal to $2\pi \delta (k+p_1+p_2+p_3).$ Therefore,

\begin{eqnarray}\label{eq30}
n_{kk'}(T) \approx \lambda^2 \frac{1}{2^4\cdot\sqrt{kk'}} \int \prod_{i=1}^3\frac{dp_i}{2\pi \, p_i} \, 2\pi\delta(k+p_1+p_2+p_3)\cdot 2\pi\delta(k'+p_1+p_2+p_3)\times \nonumber \\
\times \bigg(\sum_{\sigma,\sigma_i=\pm 1} \sigma\sigma_1\sigma_2\sigma_3\ \frac{1}{i(\sigma k+\sum_i \sigma_i p_i-i\epsilon)}\bigg)\cdot \bigg(\sum_{\sigma,\sigma_i=\pm 1} \sigma\sigma_1\sigma_2\sigma_3\ \frac{-1}{i(\sigma k'+\sum_i \sigma_i p_i+i\epsilon)}\bigg).  \label{eq:nkk1}
\end{eqnarray}
Because $k,p_1,p_2,p_3\geq0$, the equality $k+p_1+p_2+p_3=0$ can be fulfilled only when $k=p_1=p_2=p_3=0$. Hence, due to the presence of the $\delta$-functions the expressions in the round brackets are equal to zero, and $n_{kk'}=0$. Thus, again in the stationary situation we do not encounter any secular growth.

It is probably worth stressing at this point that in eq. (\ref{eq30}) we encounter the standard complication due to the infrared divergence of the massless case at $p_{1,2,3} = k = k' = 0$. To regulate the divergence one has to consider the massive case. In the latter situation the integrand in (\ref{eq:nkk1}) is proportional to:

\begin{eqnarray}
\frac{1}{\sqrt{\epsilon_k \, \epsilon_{k'}} \, \epsilon_{p_1} \, \epsilon_{p_2} \, \epsilon_{p_3}} \, \delta\bigg(\varepsilon_k+\sum_i \varepsilon_{p_i}\bigg)\delta\bigg(\varepsilon_{k'}+\sum_i \varepsilon_{p_i}\bigg)\cdot \bigg(\sum_{\sigma,\sigma_i=\pm 1} \sigma\sigma_1\sigma_2\sigma_3\ \frac{1}{i(\sigma k+\sum_i \sigma_i p_i-i\epsilon)}\bigg)\times \nonumber \\
\times \bigg(\sum_{\sigma,\sigma_i=\pm 1} \sigma\sigma_1\sigma_2\sigma_3\ \frac{-1}{i(\sigma k'+\sum_i \sigma_i p_i+i\epsilon)}\bigg).
\end{eqnarray}
As in the previous subsection, the argument of the $\delta$-function, $\varepsilon_k+\sum_i \varepsilon_{p_i}$, is never equal to $0$, so the correction, in fact, is vanishing.

\subsection{The broken world-line and world--line (\ref{worldline}) with $\beta < 1$}

In this case we use the modes from eq. (\ref{eq:7}).
Then the integral over $z$ in eq.(\ref{eq:nkk}) can be divided as follows:

$$\int d^2 z\  \theta(T-z^0)\  h_k(z) h_{p_1} (z) h_{p_2} (z) h_{p_3} (z)=\int_{t_0}^{0} dz^0 \int_{0}^{\infty} dz^1\ h_k^0(z) h_{p_1}^0 (z) h_{p_2}^0 (z) h_{p_3}^0 (z)+$$
$$+\int_{0}^{T} dz^0 \int_{-\beta z^0}^{z^0} dz^1\ h_k^\beta(z) h_{p_1}^\beta (z) h_{p_2}^\beta (z) h_{p_3}^\beta (z)+\int_{0}^{T} dz^0 \int_{z^0}^{\infty} dz^1\ h_k^0(z) h_{p_1}^0 (z) h_{p_2}^0 (z) h_{p_3}^0 (z).$$
Here $h^0$ and $h^\beta$ are defined in (\ref{h0hbeta}).

Adding the term $$\int_0^T dz^0 \int_{0}^{z^0} dz^1\ h_k^0(z) h_{p_1}^0 (z) h_{p_2}^0 (z) h_{p_3}^0 (z) $$
to the r.h.s. of the above equation and subtracting it, we obtain the contribution to $n_{kk'}$ from the previous subsection:

$$\int_{t_0}^{T} dz^0 \int_{0}^{\infty} dz^1\ h_k^0(z) h_{p_1}^0 (z) h_{p_2}^0 (z) h_{p_3}^0 (z),$$
which tends to $0$ when $T - t_0 \to +\infty$, as we have observed above.

Consequently, in the limit $T - t_0 \to +\infty$, the integral over $z$ can be simplified to:

\begin{eqnarray}
\int d^2 z\  \theta(T-z^0)\  h_k(z) h_{p_1} (z) h_{p_2} (z) h_{p_3} (z) \approx \nonumber \\ \approx \int_{0}^{T(1+\beta)} du \int_{\alpha u}^{2T-u} dv\ h_k^\beta h_{p_1}^\beta h_{p_2}^\beta h_{p_3}^\beta -\int_0^T du \int_{u}^{2T-u} dv\ h_k^0 h_{p_1}^0 h_{p_2}^0 h_{p_3}^0. \end{eqnarray}
As a result the leading contribution to $n_{kk'}$ in powers of $T$ is as follows:

\begin{eqnarray}
\label{eq:nkkbroken}
n_{kk'}\approx  (\lambda \, T)^2  \frac{(1-\alpha)^2}{2 \, \alpha^2} \int \prod_{j=1}^3 \frac{dp_j}{2 \pi \, p_j} \frac{1}{\sqrt{kk'}}\bigg[\pv \frac{i}{k+p_1+p_2+p_3} + \nonumber \\ + \frac{1+2\alpha}{\alpha}\pi\delta(k+p_1+p_2+p_3)\bigg]
\cdot \bigg[-\pv \frac{i}{k'+p_1+p_2+p_3}+\frac{1+2\alpha}{\alpha}\pi\delta(k'+p_1+p_2+p_3)\bigg].
\end{eqnarray}
Similarly to the case of the standing mirror from the previous subsection, the terms containing $\delta$-functions in this expression give zero contributions to $n_{kk'}$. So the leading contribution can be simplified to:

\begin{equation}\label{eq:nkkbrokensimple}
n_{kk'}\approx \  (\lambda \, T)^2\  \frac{(1-\alpha)^2}{2\, \alpha^2} \int \prod_{j=1}^3 \frac{dp_j}{2\pi \, p_j} \, \frac{1}{\sqrt{kk'}}\bigg[\pv \frac{1}{k+p_1+p_2+p_3}\bigg]\cdot\bigg[ \pv \frac{1}{k'+p_1+p_2+p_3}\bigg].
\end{equation}
Again we assume that the problems in this expression at $p_i = 0$ are resolved via the introduction of a small, but non--zero mass, as was explained above.

\begin{figure}[h]
\centering
\includegraphics[scale=0.26]{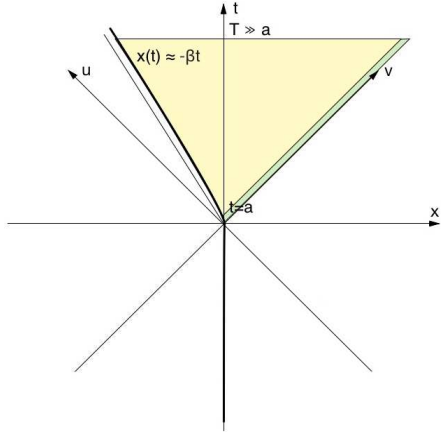}
\caption{The world-line (\ref{worldline}). In the yellow region the modes for the case of this world-line are approximately equal to those for the case of the broken world-line. In the green region they are different from each other.}
\label{fig:asymptbeta}
\end{figure}

It is probably worth stressing at this point that in the limit $T\to \infty$ the leading loop corrections for the case of the broken world-line well approximate the corrections for the case of the world--line (\ref{worldline}) with $\beta<1$.
In fact, in the latter case one should use $h_k(u,v) \propto e^{-ikv}-e^{-ikp(u)}$ instead of (\ref{eq:7}), but the two types of modes under consideration coincide with each other in the region $u<0$ and are almost equal in the region $u \gg a$ (see fig. \ref{fig:asymptbeta}). So, in the limit $T\to \infty$ the leading loop corrections in the cases of the broken world-line and of the world-line (\ref{worldline}) are approximately equal to each other.

The secular growth that we observe here is the revelation of the fact that in a non--stationary situation in a self--interacting theory the particle number density\footnote{Note that the particle number density is $\left\langle a^+_k \, a_{k'}\right\rangle \, h_k(\underline{x}) \, \bar{h}_{k'}(\underline{x}) = n_k(\underline{x}) \, \delta\left(k-k'\right)$ rather than just $\left\langle a^+_k \, a_{k'}\right\rangle$. But still we will call the latter as the particle number density just to simplify the reasoning.}, $n_{kk'}$, is changing in time. And even if $\lambda \ll 1$, for long enough time evolution of the system it happens that $\lambda T \sim 1$. This means the break down of the perturbation theory. Hence, to understand the physical consequences one needs to perform a resummation of at least the leading corrections from all loops. We come back to this point in the concluding section.

\subsection{Eternally accelerating mirror} \label{nkkexp}

Let us continue with the calculation of $n_{kk'}(T)$ for the mirror world-line (\ref{worldline}) with $\beta = 1$.
Applying the same arguments as in the previous subsection, in the limit $T - t_0 \to +\infty$ we obtain the following expression for the integral over $z$ contributing to (\ref{eq:nkk}):

\begin{equation}\begin{split}
\label{eq:simpleexp} & \int d^2 z\  \theta(T-z^0)\  h_k(z) h_{p_1} (z) h_{p_2} (z) h_{p_3} (z) = \\
& =\int_{0}^{T-x(T)} du \int_{2t_u-u}^{2T-u} dv\ h_k^\beta h_{p_1}^\beta h_{p_2}^\beta h_{p_3}^\beta -\int_0^T du \int_{u}^{2T-u} dv\ h_k^0 h_{p_1}^0 h_{p_2}^0 h_{p_3}^0. \end{split}\end{equation}
The second integral is the same as in the previous section. In the first integral it is convenient to change the integration variables as follows:

$$dt_u\bigg[1-\frac{dx(t_u)}{dt_u}\bigg]=du.$$
Then the integral in question transforms into:

$$ \int_{0}^{T} dt_u\bigg[2-e^{-t_u/a}\bigg]\  e^{-i(k+p_1+p_2+p_3-i\epsilon)a(1-e^{-t_u/a})} \bigg[\frac{1}{i(k+p_1+p_2+p_3-i\epsilon)}+$$
$$+2T-2t_u-\frac{1}{i(k+p_1+p_2-i\epsilon)}-\frac{1}{i(p_3-i\epsilon)}+...\bigg].$$
Where the dots stand for the subleading terms in the limit $T\to +\infty$.

\begin{figure}[h]
\centering
\includegraphics[scale=0.26]{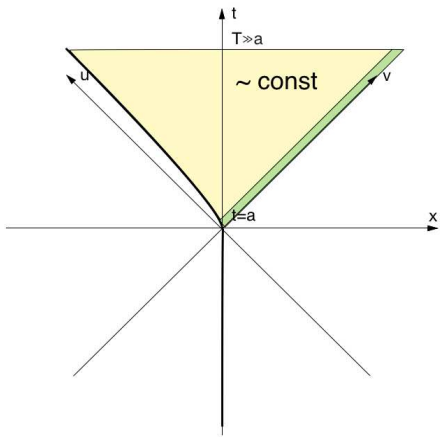}
\caption{The region of integration in the integral (\ref{eq:eqeq}). In the yellow region the reflected wave is almost constant. In the green region it behaves like a plane wave.}
\label{fig:6}
\end{figure}

If the world-line had the form $x(t)=-\beta t+a(1-e^{-\beta t/a})$ for $t>0$ and $\beta < 1$, then, as a result of the integration in the last equation, we would obtain $\delta$-functions due to the proportionality of the argument of the exponent to $t_u$. Indeed, in such a case we would have had $t_u+x(t_u)=t_u(1-\beta)+a(1-e^{-\beta t/a}) \sim t_u$, as $t \to +\infty$. When $\beta = 1$, however, the linear term in $t_u$ in the last expression vanishes and the situation is different. As a result, the leading term is as follows:

\begin{equation}
\label{eq:halfexp}
\int d^2 z\  \theta(T-z^0)\  h_k(z) h_{p_1} (z) h_{p_2} (z) h_{p_3} (z) \propto T^2 e^{-i(k+p_1+p_2+p_3-i\epsilon)a}+O(T).\end{equation}
Finally, the expression for $n_{kk'}$ to the leading order in powers of $T$ has the following form:

\begin{equation}
\label{eq:nkkexp}
n_{kk'}\propto \lambda^2 T^4\  \frac{e^{-i(k-k'-i\epsilon)a}}{ \sqrt{kk'}}\cdot  \int \prod_{j=1}^3 \frac{dp_j}{2\pi p_j}.
\end{equation}
This result can be easily understood. To calculate $n_{kk'}$, one has to take the following integral over $z$:

\begin{equation}
\label{eq:eqeq}\int_{0}^{T} dz^0 \int_{x(z^0)}^{z^0} dz\ h_k^\beta(z) h_{p_1}^\beta (z) h_{p_2}^\beta (z) h_{p_3}^\beta (z),\end{equation}
where $h_p^\beta(z)\propto e^{-ipv}-e^{-ipa(1-e^{-t_u/a})}$. The reflected wave behaves as a constant in the region $u\gg a$, and as an exponent in the region $u\ll a$:

\begin{eqnarray}\label{UVmodes}
e^{-ipa(1-e^{-t_u/a})} \sim \begin{cases}
e^{-ipa}\ \ \ u\gg a\\
e^{-ipt_u}\ \   u\ll a.
\end{cases}
\end{eqnarray}
As a result, due to the behaviour of reflected waves in the limit when $T\gg a$, one obtains the expression, which is proportional to the volume of the region of integration, $\propto T^2$, (see fig. \ref{fig:6}). Thus, $n_{kk'}$ is proportional to $T^2 \cdot T^2=T^4$.

Note also that when there is a violation of the Lorentz invariance one can have a non--vanishing $n_{kk'}$ for the case when $k\neq k'$. Then such a $n_{kk'}$ can be complex. It just a consequence of the fact that we are dealing here with a spatially non--homogeneous situation.

Finally, the integrals over $p_i$ in eq. (\ref{eq:nkkexp}) contain ultraviolet divergences\footnote{They are present on top of the infrared divergences, which are typical in such integrals for the massless field and can be regularized by introducing a small mass, as we have explained above.}. To regulate these divergences one has to consider a non-perfect mirror of some sort. In fact, the divergences come from the fact that the mode functions is (\ref{UVmodes}) behave as constants when $u \gg a$. In the case of non--perfect mirror the mode functions will depend on space and time coordinates in the limit in question. This dependence should regulate the divergence under consideration. Just to give an idea how it should work, consider modes, which have the following from

$$
h_k(u,v) \propto \alpha(k) \, e^{- i \, k \, v} - \beta(k) \, e^{- i \, k \, p(u)}.
$$
Somewhat similar form they should have in the presence of a non--ideal mirror. In fact, imagine that $\alpha(k) \sim \beta(k) \sim 1$ for $|k| < \Lambda$ and $\beta(k) = 0$ for $|k| > \Lambda$, where $\Lambda$ is some energy scale. The behavior of the modes under consideration just means that there is a mirror which reflects all the modes with $|k| < \Lambda$ and is transparent for those with $|k| > \Lambda$.

Now, if one will recalculate the loop corrections with the use of such modes, then instead of (\ref{eq:nkkexp}) he would obtain:

$$
n_{kk'}\propto \lambda^2 T^4\, \beta(k) \bar{\beta}(k')  \frac{e^{-i(k-k'-i\epsilon)a}}{ \sqrt{kk'}}\cdot  \int \prod_{j=1}^3 \frac{dp_j \, \left|\beta_k(p_j)\right|^2}{2\pi p_j},
$$
which obviously does not have any UV divergence. But, in any case, a more rigorous calculation will be done elsewhere.

\section{The two-loop corrections to $\kappa_{kk'}$}

The calculation of $\kappa_{kk'}$ from eq. (\ref{eq:kappakk}) is very similar to the one of $n_{kk'}$. Hence, in this section we just concisely list the results.

\subsection{The mirror at rest}

We have to calculate the following quantity, which is a part of the contribution to $\kappa_{kk'}$:

\begin{equation}\label{eq:partkappa}
\int \bigg(\prod_{i=1}^3\frac{dp_i}{2\pi}\bigg) \int_{t_0}^T dz^0 \int_{0}^{\infty} dz^1\  \bar h_k(z^0,z) \prod_{i=1}^3 h_{p_i}(z^0,z) \int_{t_0}^{z^0} dw^0 \int_{0}^\infty dw^1\  \bar h_{k'}(w^0,w)\prod_{i=1}^3 \bar h_{p_i} (w^0,w).\end{equation}
After the substitution of the explicit from of the mode functions the correction to $\kappa_{kk'}$ can be rewritten as follows:

\begin{equation}
\label{eq:kapparest}
\begin{split}
\kappa_{kk'}=&-\frac{\pi^2}{2^4 \, \sqrt{kk'}}\int \bigg(\prod_{i=1}^3\frac{dp_i}{2\pi p_i}\bigg)\  \cdot\frac{1}{i(p_1+p_2+p_3+(k'-k)/2-i\epsilon)}\cdot4\pi\delta(k+k')\times\\
&\times \bigg[\sum_{\sigma,\sigma_i=\pm1} \sigma\sigma_1\sigma_2\sigma_3\  \delta (\sigma_1 p_1+\sigma_2 p_2+\sigma_3 p_3)\bigg]^2.\end{split}\end{equation}\
The sum in square brackets is equal to zero. Then, as is expected, in this case $\kappa_{kk'}=0$. The infrared divergences of the last expression at $p_i = 0$ are taken care of in the same manner as in the previous subsections.

\subsection{The broken world-line}

The loop correction to $\kappa_{kk'}$ for the broken world-line is as follows:

\begin{equation}
\begin{split}
\label{eq:kappabroken}
\kappa_{kk'}\approx\   & \lambda^2 T^2\frac{1}{\sqrt{kk'}} \frac{\alpha-1}{\alpha} \int \bigg(\prod_{i=1}^3\frac{dp_i}{2\pi p_i}\bigg) \, \bigg\{ \frac{1}{i(k'+p_1+p_2+p_3)}\bigg[b_\alpha\cdot \pv\frac{1}{-k+p_1+p_2+p_3}+ \\
&+ c_\alpha\cdot\delta(-k+p_1+p_2+p_3) \bigg]-d_\alpha\cdot \delta(k'+p_1+p_2+p_3)\bigg[ \pv \frac{i}{k+p_1+p_2+p_3}+ \\
&+ \frac{1+2\alpha}{\alpha}\pi\delta(k+p_1+p_2+p_3)\bigg]\bigg\}  +\big\{k\leftrightarrow k'\big\},
\end{split}
\end{equation}
where $d_\alpha$, $b_\alpha$ and $c_\alpha$ are some momentum and time independent coefficients, which, however, depend on $\alpha$.
Their exact from is not very relevant for the estimates of the leading terms that we consider here.

After dropping off the last term, proportional to $\delta(k'+\sum p_i)$ for the same reason as it was done in the previous section, one obtains that:

\begin{equation}\begin{split} \kappa_{kk'}\approx\  & \lambda^2 T^2\frac{1}{\sqrt{kk'}} \frac{\alpha-1}{\alpha} \int \bigg(\prod_{i=1}^3\frac{dp_i}{2\pi p_i}\bigg) \cdot \frac{1}{i(k'+p_1+p_2+p_3)}\bigg[b_\alpha\cdot \pv \frac{1}{-k+p_1+p_2+p_3}+\\
& +c_\alpha\cdot\delta(-k+p_1+p_2+p_3) \bigg]+\big\{k\leftrightarrow k'\big\}.\end{split}
\end{equation}
This result for $\kappa_{kk'}$ is not zero, as is expected in the non--stationary situation in question, because modes that diagonalize the linear combination $H - \beta(t) \, P$ in the past and future infinity are not equal to each other. In fact, as we have explained in the Introduction, the secular growth of $\kappa_{kk'}$ in a non--stationary situation is just a signal of the fact that the ground state of the free Hamiltonian at past infinity does not remain to be such a state at future infinity.

\subsection{Eternally accelerating mirror}

In this case the correction to $\kappa_{kk'}$ in the leading approximation, as $T\to \infty$, is as follows:

\begin{equation}
 \label{eq:kappaexp}
 \kappa_{kk'}\propto \lambda^2 T^4\  \frac{e^{i(k+k')a}}{ \sqrt{kk'}}\cdot  \int \prod_{j=1}^3 \frac{dp_j}{2\pi \, p_j}.
 \end{equation}
This correction has the same UV divergences as $n_{kk'}$ in the same situation. Again this problem is resolved along the same lines as were proposed in the subsection 3.4.

\section{One-loop corrections to the Keldysh propagator}

We continue with the consideration of the one loop bubble diagram contribution to the Keldysh propagator:

\begin{equation}
\label{eq:diagoneloop}
\Delta G^{K}_{xy}=\sum_{\sigma,\sigma_1=\pm}
\begin{tikzpicture}

\begin{feynman}

 \vertex[label=${x, \sigma}$] (x);
    \vertex[label=${z, \sigma_1 }\ \ \ \ \ \ \ \ \ \ \ $] [right=2.5cm of x] (z);
    \vertex[label=${y, \sigma}$] [right=2.5cm of z] (y);

    \diagram* {

      (x) -- [plain] (z) -- [plain] (y)
      (z) -- [plain, out=130, in=50, loop, min distance=3cm] (z)

    };

   \end{feynman}
 \end{tikzpicture}
    \end{equation}
Using the expressions of the propagators through the mode functions (\ref{eq:throughmodes}), one can obtain the following contributions to $n_{kk'}$ and $\kappa_{kk'}$ coming from such a loop correction in the limit $\displaystyle \frac{x^0+y^0}{2}= T  \to + \infty,$ and $\big|x^0-y^0\big|=\operatorname{const}$:

$$
n_{kk'}\propto \lambda \int d^2 z\ \bigg[\theta(x^0-z^0)\theta(z^0-y^0)-\theta(z^0-x^0)\theta(y^0-z^0)\bigg]\ h_k(z)\bar h_{k'}(z)  \int \frac{dp}{2\pi}\ h_p(z)\bar h_p(z)\bigg|_{x^0\approx y^0\approx T} \approx 0.
$$
(Thus, for any type of the mirror motion we obtain that the contribution to $n_{kk'}$ coming from the one loop is always zero.) At the same time:

\begin{eqnarray}\label{eq:oneloopkappa}
\kappa_{kk'} \propto \lambda \int d^2 z\  \theta(T-z^0)\  \bar h_k(z) \bar h_{k'} (z) \int \frac{dp}{2\pi}\ h_p (z)\bar h_{p} (z) = \nonumber \\ = \lambda \int \frac{dp}{2\pi}\ \int d^2 z\  \theta(T-z^0)\  \bar h_k(z) \bar h_{k'} (z) h_p (z)\bar h_{p} (z).
\end{eqnarray}
and we have already calculated the last integral in the section \ref{twoloopnkk} above. Performing the following change of variables:

$$k \to -k,\ p_1 \to -k',\ p_2 \to p,\ p_3 \to -p,$$
one obtains that for the mirror at rest:

\begin{equation}\kappa_{kk'}\propto \lambda \int \frac{dp}{2\pi} \frac{1}{\sqrt{kk'p^2}}\cdot 2\pi\delta(k+k')\cdot\bigg(\sum_{\sigma,\sigma', \sigma_i=\pm 1} \sigma\sigma'\sigma_1\sigma_2\ \frac{1}{i(-\sigma k-\sigma' k'+\sigma_1 p-\sigma_2 p-i\epsilon)}\bigg) \approx 0.
\end{equation}
At the same time, for the broken world--line:

\begin{equation}
\begin{split}
\kappa_{kk'} \approx -\lambda\ \frac{2T(1-\alpha)}{\alpha} \int \frac{dp}{2\pi} \frac{1}{\sqrt{2^4\cdot kk'p^2}}\cdot \bigg[- \pv \frac{i}{k+k'}+\frac{1+2\alpha}{\alpha}\pi\delta(k+k')\bigg] + O(1).\end{split}
\end{equation}
And, finally, for the mirror, approaching the speed of light we find that:

\begin{equation}
\kappa_{kk'} \propto \lambda T^2 \int \frac{dp}{2\pi} \frac{1}{\sqrt{kk'p^2}}\cdot e^{-i(-k-k'-i\epsilon)a}+O(T)\end{equation}
Here again we encounter the same sort of problems as above for $n_{kk'}$ and resolve them in the same way.
Thus, there is also a secular growth in one loop bubble diagrams. However, in the ressumation (inside Dyson--Schwinger equations) bubble diagrams contribute to a time--dependent mass renormalization rather than to the change of the state of the theory.

 \section{Conclusions and acknowledgements}

We have shown that two--point Wightman and/or Keldysh correlation functions in the theory under consideration grow with time. This growth reveals itself through the time dependence of the particle number density $\left\langle a^+_k \, a_{k'} \right\rangle$ and of the anomalous quantum average $\left\langle a_k \, a_{k'} \right\rangle$ in the self--interacting theory.

We find the secular growth in question in the first loop bubble diagram and in the second loop sunset diagram. In the first loop we find that the anomalous quantum average grows as $\left\langle a_k \, a_{k'} \right\rangle \sim \lambda T$ for the broken world line of the mirror and $\left\langle a_k \, a_{k'} \right\rangle \sim \lambda T^2$ for the eternally accelerating mirror. At the same time the number density, $\left\langle a^+_k \, a_{k'} \right\rangle$, does not change. (Here $T$ is the average time of the two points in the correlation function.) In the second loop both the particle number density, $\left\langle a^+_k \, a_{k'} \right\rangle$, and the anomalous quantum average, $\left\langle a_k \, a_{k'} \right\rangle$, grow as $\left(\lambda T\right)^2$ for the broken world line and as $\left(\lambda T^2\right)^2$ for the case of the eternal acceleration.

Thus, even if $\lambda \ll 1$ for big enough $T$ we have a break down of the perturbation theory.
Hence, to understand the physical essence of the phenomenon under consideration we need to resum at least the leading secularly growing corrections from all loops. The situation is quite similar in spirit to the resummation of the leading UV divergences $\lambda^2 \log \Lambda$ with the use of the renormalization group. The difference, however, is that now we have to deal with something which is IR and, hence, non--local. To do the resummation one has to solve the system of the Dyson--Schwinger equations for the propagators, self--energies and vertexes. The result of the resummation will provide the correct time dependence of the particle number density and of the anomalous quantum average, i.e. the proper change of the state of the quantum field theory under the effect of the mirror motion. The result can depend on the initial conditions. In the present paper we have considered vacuum state before the start of the mirror motion, but, in principle, one can extend all these considerations to the case of non--zero initial particle density.

The problem, however, is that in the present situation we do not observe the standard kinetic behaviour of the correlation functions. In the latter one the particle density and anomalous average show the linear growth of such form as $\lambda^2 T$, which is coming from the sunset diagrams.

To illustrate what is going on here consider flat space with initial non--Planckian particle density. Then, if $\lambda = 0$ this density matrix does not change in time. However, if one turns on self--interactions, $\lambda \neq 0$, the particle density starts to change in time due to scattering of particles from different levels. If in the calculation of the rate of such a dynamics one neglects the change of the density (i.e. one calculates the collision integral with the use of the initial density rather than with the use of the exact one \cite{LL}, \cite{Kamenev}), then he observes a linear secular growth. After the resummation of the leading divergences from all loops, via the reduction of the Dyson--Schwinger equation to the Boltzman one, one finds the proper time dependence of the particle density. Obviously in the future it relaxes to the Planckian distribution with a temperature which depends on the initial conditions. In fact, the Planckian distribution annihilates the collision integral.

Please note the striking phenomenon: if $\lambda$ is always zero the density remains always the same and is non--Planckian; If, however, one adiabatically turns on $\lambda$, keeps it non--zero for long enough time and, then, puts back to zero adiabatically, the density changes to the Planckian one, which then can stay unchanged indefinitely, if there is no any external influence on the system. This is the non--perturbative phenomenon.

In the situation considered in this paper, however, we have several complications. The first one is that on top of the particle density one has to deal with the anomalous average. In \cite{Akhmedov:2013vka} it is explained how to work in such a situation, if both $\left\langle a_k \, a_{k'} \right\rangle$ and $\left\langle a^+_k \, a_{k'} \right\rangle$ grow linearly in time. Second complication is that here we see a non--linear growth in time.

And, finally, the third complication is that in the standard kinetic situation (that have been studied in the literature) there is only growth of two--point functions, while in the situation that we study in the present paper there is also a the secular growth in the vertexes, i.e. in the multiple point correlations. In fact, consider the following correction to the four-point correlation function in the limit when $x_0^1, x_0^2, x_0^3, x_0^4 \sim T \to +\infty$, where the distances $|x_0^i-x_0^j|$ for all $i< j$ are kept fixed:

\begin{equation}
\Delta G^{++--}_{x_1x_2x_3x_4}=\sum_{\sigma_i=\pm}
\begin{gathered}
\begin{tikzpicture}
\begin{feynman}

 \vertex[label=${x_1, +}$] (x1);
    \vertex[label=0:$\ {z, \sigma_1 }$] [below right=1cm and 1.5cm of x1] (z);
    \vertex[label=${x_2, +}\ \ \ $] [below left=1cm and 1.5cm of z] (x2);
    \vertex[label=180:${w, \sigma_2}\ $] [right=4cm of z] (w) ;
    \vertex[label=${x_3, - }$] [above right=1cm and 1.5cm of w] (x3);
    \vertex[label=$\ \ \ \ \ {x_4, - }$] [below right=1cm and 1.5cm of w] (x4);

    \diagram* {

      (x1) -- [plain] (z) -- [plain] (x2)
      (z) -- [plain, out=55, in=125] (w)
      (z) -- [plain, out=-55, in=-125] (w)
      (x3) -- [plain] (w) -- [plain] (x4)

    };

   \end{feynman}
 \end{tikzpicture}
 \end{gathered}
\end{equation}
The sum of these diagrams is proportional to

$$\lambda^2 \int d^2 z \int d^2 w\ \bigg\{ G^{+-}_{x_1 z} G^{+-}_{x_2 z} \big[G^{-+}_{zw}\big]^2 G^{+-}_{w x_3} G^{+-}_{w x_4}-G^{+-}_{x_1 z} G^{+-}_{x_2 z} \big[G^{--}_{zw}\big]^2 G^{--}_{w x_3} G^{--}_{w x_4}+$$
$$+G^{++}_{x_1 z} G^{++}_{x_2 z} \big[G^{+-}_{zw}\big]^2 G^{--}_{w x_3} G^{--}_{w x_4}-G^{++}_{x_1 z} G^{++}_{x_2 z} \big[G^{++}_{zw}\big]^2 G^{+-}_{w x_3} G^{+-}_{w x_4}\bigg\}.$$
In analogy with eq. (\ref{GK}), we represent this correction in the following form:

$$ \int \bigg(\prod_{i=1}^4 \frac{dk_i}{2\pi}\bigg)\ \bigg[n_{k_1k_2k_3k_4}^{(1)}\cdot\  h_{k_1}(x_1)h_{k_2}(x_2)\bar h_{k_3}(x_3)\bar h_{k_4}(x_4)+n_{k_1k_2k_3k_4}^{(2)}\cdot\  h_{k_1}(x_1)\bar h_{k_2}(x_2)\bar h_{k_3}(x_3)\bar h_{k_4}(x_4)+\ ...\bigg]. $$
The last sum includes all possible terms with products of $h_{k_i}(x_i)$ and $\bar h_{k_j}(x_j)$. Let us calculate e.g. the function $n_{k_1k_2k_3k_4}^{(1)}$ in the approximation under consideration:

\begin{equation}
\label{eq:nkkkk}
\begin{split}
n_{k_1k_2k_3k_4}^{(1)}& \approx  \int \frac{dp}{2\pi} \int \frac{dp'}{2\pi}\int d^2z\ \bar h_{k_1}(z)\bar h_{k_2}(z) h_p(z)h_{p'}(z) \int d^2w\  \bar h_p(w)\bar h_{p'}(w)h_{k_3}(w) h_{k_4}(w)\times \\
& \times\bigg[\theta(z^0-T)\theta(w^0-T)-\theta(z^0-w^0)\theta(w^0-T)-\theta(w^0-z^0)\theta(z^0-T)\bigg]+\\
& +\int \frac{dp}{2\pi} \int \frac{dp'}{2\pi}\int d^2z\ \bar h_{k_1}(z)\bar h_{k_2}(z) \bar h_p(z) \bar h_{p'}(z) \int d^2w\ h_p(w)h_{p'}(w)h_{k_3}(w) h_{k_4}(w)\times\\
& \times\bigg[1-\theta(w^0-z^0)\theta(w^0-T)-\theta(z^0-w^0)\theta(z^0-T)\bigg].
\end{split}
\end{equation}
Simplifying the factor, containing $\theta$--functions, from the first term above, we obtain that:

\begin{equation}
\begin{split}
& \theta(w^0-T)\big[\theta(z^0-T)-\theta(z^0-w^0)\big]-\theta(w^0-z^0)\theta(z^0-T)=\\
&=\theta(w^0-T)\theta(z^0-T)\theta(w^0-z^0)-\theta(w^0-z^0)\theta(z^0-T)=\\
&=-\theta(T-w^0)\theta(w^0-z^0)\theta(z^0-T)=0.
\end{split}
\end{equation}
Consequently, only the second term in eq. (\ref{eq:nkkkk}) contributes to $n_{k_1k_2k_3k_4}^{(1)}$, which, after a further simplification, becomes as follows:

\begin{equation}
\begin{split}
n_{k_1k_2k_3k_4}^{(1)}& \approx \int \frac{dp}{2\pi} \int \frac{dp'}{2\pi}\int d^2z\ \bar h_{k_1}(z)\bar h_{k_2}(z) \bar h_p(z) \bar h_{p'}(z) \int d^2w\ h_p(w)h_{p'}(w)h_{k_3}(w) h_{k_4}(w)\times\\
& \times\bigg[\theta(T-w^0)\theta(w^0-z^0)+\theta(T-z^0)\theta(z^0-w^0)\bigg].
\end{split}
\end{equation}
The integrals that appear here are very similar to those, which we have already calculated above. They grow with time as some power of $T$. Similarly, every term in the last equation grows with time as $T^2$ in the case of the broken world-line with $\beta<1$ or as $T^4$ in the case of the eternally accelerating mirror. Thus, the correction to the four-point correlation function coming from $n_{k_1k_2k_3k_4}^{(1)}$ is proportional to $\lambda^2T^2$ or $\lambda^2T^4$, depending on the mirror world-line. The situation is similar to the one encountered for the light and massless scalars in de Sitter space \cite{Akhmedov:2017ooy}.

Due to the presence of the secular growth in the higher point correlation functions and due its non--linear dependence on time in the lowest loop level the system of Dyson--Schwinger equations does not reduce to a Boltzman kinetic equation. Now in the leading infrared limit one has to solve a coupled system of equations simultaneously for propagators and vertexes. We do not yet know how to approach such a problem.

We would like to acknowledge discussions with L.Astrakhantsev, F.Bascone, A.Diatlyk, D.Glavan, U.Moschella and F.Popov. We would like to thank Albert Einstein Institute, Golm, Germany and Universit\`a degli Studi dell'Insubria for the hospitality during the final stage of the work on this subject. The work of ETA was done under the partial support of the RFBR grant 15-01-99504. Our work is done under the financial support of the state grant Goszadanie 3.9904.2017/BCh.

\end{document}